\documentclass[12pt]{spieman}

\usepackage{amsmath,amsfonts,amssymb}
\usepackage{graphicx}
\usepackage{setspace}
\usepackage{tocloft}
\usepackage{float}
\usepackage{booktabs}
\usepackage{siunitx}
\usepackage{hyperref}
\usepackage{placeins}
\usepackage[mathlines]{lineno}
\modulolinenumbers[1]

\cftpagenumbersoff{figure}
\cftpagenumbersoff{table}

\title{AI-based single-shot structured-light depth reconstruction for real-time laparoscopic surgical guidance}

\author[a,b,*,\dag]{Wayne Wonseok Rodgers}
\author[b,\dag]{Xiangyi Le}
\author[c]{Seonghoon Jang}
\author[a]{Shuwen Wei}
\author[b,c]{Justin Opfermann}
\author[b,c]{Michael Kam}
\author[b,c]{Axel Krieger}
\author[a,b,d]{Jin U.\ Kang}

\affil[a]{Johns Hopkins University, Department of Electrical and Computer Engineering, Baltimore, Maryland, United States}
\affil[b]{Johns Hopkins University, Laboratory for Computational Sensing and Robotics, Baltimore, Maryland, United States}
\affil[c]{Johns Hopkins University, Department of Mechanical Engineering, Baltimore, Maryland, United States}
\affil[d]{Johns Hopkins University, School of Medicine, Baltimore, Maryland, United States}

\begin{document}
\maketitle

\noindent Corresponding author: Wayne W. Rodgers, E-mail:
\href{mailto:wrodger2@jh.edu}{wrodger2@jh.edu}

\begin{abstract}
\noindent\textbf{Significance:} Accurate intraoperative depth perception is a prerequisite for autonomous and semi-autonomous robotic laparoscopic surgery. Conventional fringe projection profilometry (FPP) can achieve millimeter-scale accuracy but commonly relies on multi-shot acquisition, digital-micromirror-device (DMD) projection, and projector-camera synchronization, which complicates integration into compact laparoscopic surgical platforms.

\noindent\textbf{Aim:} To develop and characterize a synchronization-free, single-shot depth-sensing platform that replaces active DMD projection with a passive binary mask illuminated by a light-emitting diode (LED), and reconstructs dense depth from a single frame using a vector-quantized variational autoencoder (VQ-VAE) prior coupled with a custom U-Net depth head.

\noindent\textbf{Approach:} A compact LED/binary-mask projection module was designed and coupled to one of the channels of a dual-channel laparoscope (Intuitive). The remaining channel was used to image the fringe-illuminated target. A 3D vision camera (Zivid, Norway) was used to acquire $722$ reference images of a phantom target. Corresponding images using the proposed single-shot structured light endoscopic camera (SSLE) was acquired and registered. Depth measurements from both laterally mounted Zivid camera and SSLE to generate depth map. The Zivid depth maps were then reprojected into the SSLE image frame to provide reference depth maps for supervised training and evaluation. The VQ-VAE encodes the pattern-distorted image into a discrete latent representation, and a U-Net operating in the latent space predicts the corresponding depth map without requiring a separate mask-prediction branch for depth generation.

\noindent\textbf{Results:} On the SSLE--Zivid-paired binary-pattern phantom dataset, evaluated using a fixed train/validation/test split, the proposed VQ-VAE + U-Net model achieved an MAE of $3.70$~mm, AbsRel of $0.0326$, accuracy $\delta{=}1.1$ of $0.962$, and accuracy $\delta{=}1.1^2$ of $0.970$. The proposed method achieved lower MAE than the dual U-Net MaskNet + DepthNet baseline, while the baseline achieved slightly higher threshold accuracy. Compared with off-the-shelf monocular depth models, the proposed method improved MAE, AbsRel, and threshold accuracy, demonstrating the benefit of task-specific training for calibrated endoscopic depth reconstruction. The proposed pipeline operated at $26.0$~Hz over $301$ consecutive frames on an NVIDIA A100 GPU.

\noindent\textbf{Conclusions:} A compact LED-illuminated binary-pattern single-shot surface profilometer system combined with latent-space depth reconstruction provides a practical route toward synchronization-free video-rate depth perception for robotic laparoscopy. The results demonstrate Zivid-referenced reconstruction on a phantom dataset with no explicit segmentation stage in the depth-generation path, while highlighting the importance of dataset size and SSLE--Zivid calibration accuracy for cross-camera supervised training.
\end{abstract}


\noindent This manuscript has been submitted to the Journal of Biomedical Optics for consideration.\\

\section{Introduction}

Autonomous laparoscopic robotic surgery requires a dense, accurate, real-time depth map of the operative field, both for surgeon visualization and for closed-loop autonomous tool control.\cite{Saeidi2022SciRobot,Kam2019MICCAI} Learning-based monocular depth estimation has been explored in endoscopy without projected patterns,\cite{Liu2020TMI} but monocular methods recover depth only up to an unknown scale unless metric supervision or external calibration is available, limiting their direct use for closed-loop tool control. Complementary real-time optical modalities, such as OCT-based tissue classification for autonomous intestinal anastomosis,\cite{Wang2024BOE} reflect a broader trend toward compact optical perception on robotic surgical platforms. Among metric 3D techniques, fringe projection profilometry (FPP) has long been a workhorse for surface metrology because of its high accuracy, large field of view, dense reconstruction, and real-time imaging potential,\cite{zhang2010review,Geng2011AOP,Nguyen2015AO} and it has been demonstrated in laparoscopic and endoscopic settings.\cite{Ackerman2002SPIE,Clancy2011BOE,Le2018JBO,Wei2022JOSAA} However, conventional FPP implementations rely on multi-frequency or temporally phase-unwrapped sinusoidal patterns, spatial-light-modulator or digital-micromirror-device (DMD) projection, and precise projector--camera synchronization.\cite{Saldner1997OE,zuo2018review} Each of these requirements complicates clinical translation: multi-shot acquisition is vulnerable to physiological motion from respiration, cardiac pulsation, and tool--tissue interaction; DMD-based projectors are bulky and thermally demanding; and hardware synchronization adds complexity and potential failure modes that are poorly suited to compact clinical systems. Accurate FPP reconstruction also depends on careful geometric calibration of the camera--projector configuration, particularly when the system geometry is flexible or arbitrarily arranged.\cite{Du2007OL,Vo2012OE,Vo2011OE}

Recent deep-learning approaches to fringe analysis have removed the multi-shot requirement, first for single-frame phase demodulation\cite{Feng2019AP} and subsequently for direct depth regression from a single fringe image.\cite{Nguyen2021Photonics,Wang2021OE,Zuo2025JBO} Zuo et al.\cite{Zuo2025JBO} demonstrated an endoscopic single-shot FPP system using a dual-channel laparoscope and a two-network pipeline, in which a segmentation network (MaskNet) and a depth-regression network (DepthNet) are combined multiplicatively. That system achieved a mean absolute error (MAE) of $2.28$~mm with binary patterns ($2.38$~mm with sinusoidal patterns) on experimental phantom data at $20$ frames-per-second acquisition, under a supervision protocol in which the ground truth was derived from the conventional multi-shot FPP algorithm itself (see Sec.~\ref{sec:Discussion}), establishing that binary patterns are at least competitive with sinusoidal patterns once a learned model is used for depth reconstruction. Several aspects of that design nonetheless leave room for improvement. Although the binary-pattern result opened the door to replacing the active projector with a passive mask, the optical chain still relied on a DMD projector with its accompanying driver electronics, collimation optics, and thermal load. The FPP-derived ground truth tied the learned model to the accuracy and noise characteristics of the supervisory pipeline. Finally, the MaskNet + DepthNet pipeline showed substantial frame-to-frame variability in inference latency, which complicates integration into a synchronous robotic control loop. Notably, our prior work identified the first two of these as directions for future development: an LED source with a passive binary mask, and ground truth obtained independently of the FPP algorithm.\cite{Zuo2025JBO} The present work realizes both, and addresses the third.

In this work, we present a single-shot structured light endoscopic camera (SSLE) system that addresses each of these limitations. First, a compact projection module replaces the DMD with an LED-illuminated passive binary mask that fits within the illumination channel of a dual-channel laparoscope. Second, a learned depth-reconstruction pipeline employs a vector-quantized variational autoencoder (VQ-VAE)\cite{vanDenOord2017VQVAE} to produce a regularized discrete latent representation of the pattern-distorted image, and a multi-residual decoder U-Net\cite{Ronneberger2015UNet,Zuo2025JBO} operating on this latent to predict the corresponding depth representation, which is decoded into the final depth map. Third, the training and evaluation protocol obtains depth ground truth from an independent Zivid structured-light camera reprojected into the endoscopic camera frame, decoupling the learned model from the conventional FPP algorithm. The system is targeted at compact robotic platforms (\textit{e.g.}, UR5-style manipulators) and is designed to operate synchronization-free at the camera frame rate. On a 722-pair SSLE--Zivid phantom dataset, the proposed pipeline achieves an MAE of $3.70$~mm with a $\delta = 1.1$ accuracy of $0.962$, and sustains a stable $26.0$~Hz inference rate on an NVIDIA A100 GPU. A preliminary version of the hardware design was presented at SPIE Photonics West 2026;\cite{Rodgers2026SPIE} the present work extends that conference report with the VQ-VAE-regularized reconstruction network, the Zivid-referenced ground-truth pipeline, and the associated quantitative characterization.

\section{Methods}
\label{sec:methods}

\subsection{System overview and mechanical integration}
\label{sec:system-overview}

The proposed SSLE platform consists of a passive binary mask illuminated by an LED, a pair of relay lenses, and a dual-channel laparoscope whose illumination channel is used to project the binary pattern and the remaining channel is used to image the scene. Because the mask is static and the source is a DC-driven LED, there is no synchronization link between the projection and the camera: any frame captured by the camera carries a valid pattern.

\begin{figure}[H]
\centering
\includegraphics[width=0.95\linewidth]{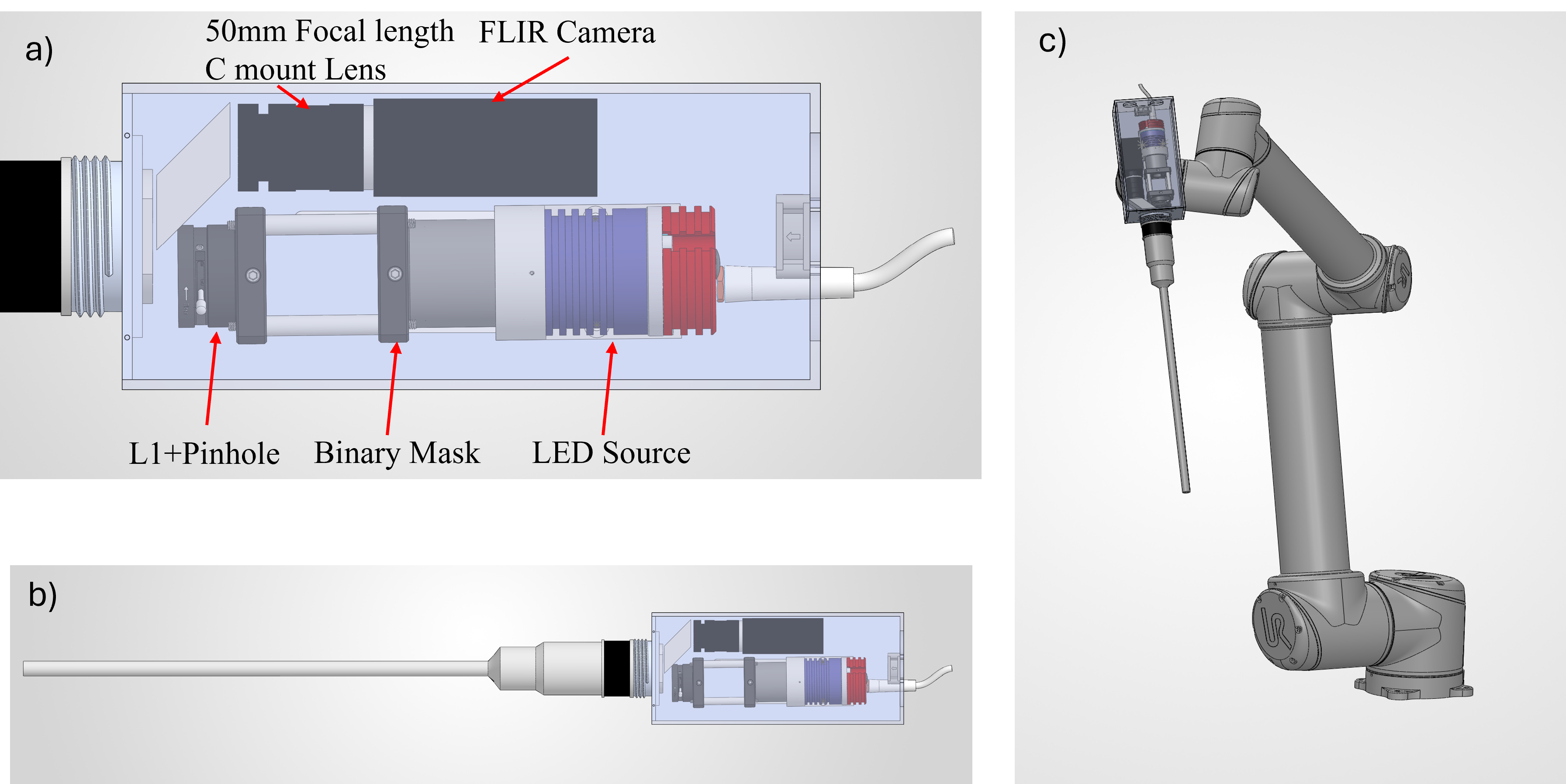}
\caption{Mechanical integration of the LED-illuminated binary-pattern projection module. (a) Cutaway of the compact projection unit showing the LED source (right, red heatsink), relay optics, and passive binary mask (left, angled). (b) The projection unit mated to the illumination channel of a dual-channel laparoscope; the imaging channel is coaxial and unobstructed. (c) Integration onto the camera arm of a UR5-style robotic platform, illustrating compatibility with the surgical manipulator envelope.}
\label{fig:cad}
\end{figure}

Figure~\ref{fig:cad} summarizes the mechanical integration. The compact projection unit in Fig.~\ref{fig:cad}(a) houses the LED, mask, and relay inside a sealed enclosure designed to thread directly onto the illumination port of the laparoscope. The cutaway in Fig.~\ref{fig:cad}(a) and the side view with the endoscope shaft in Fig.~\ref{fig:cad}(b) show that the projection module occupies a volume comparable to a standard light-guide connector and that the endoscope imaging channel remains unobstructed. Figure~\ref{fig:cad}(c) illustrates the integration of the assembled laparoscope onto the camera arm of a UR5-style robotic platform, which we use here as a representative surgical manipulator; the bounding envelope of the projection module does not interfere with the articulation envelopes of the tool arms.

The binary mask encodes a series of alternating opaque and transmissive stripes with a target duty cycle of $1{:}1$. Two fabrication routes were evaluated in prior work: a low-cost 3D-printed mask and a commercial photolithographic chrome-on-glass mask. The 3D-printed mask achieved the target duty cycle and is used throughout this study to facilitate rapid iteration on pattern frequency and mask geometry.

\subsection{Optical setup and bench realization}
\label{sec:optical-setup}

\begin{figure}[H]
\centering
\includegraphics[width=0.95\linewidth]{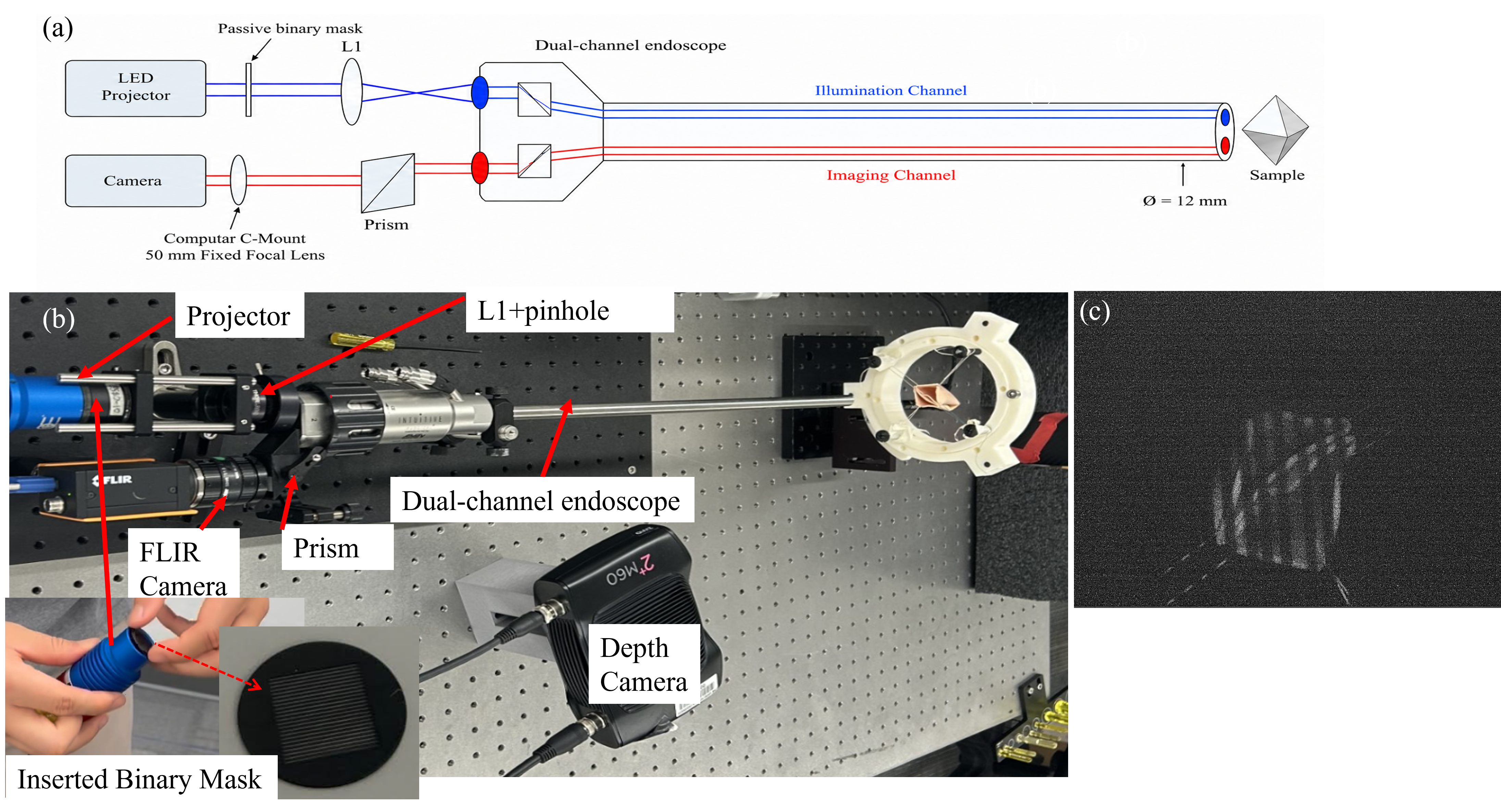}
\caption{Optical configuration and bench realization of the binary-pattern single-shot platform. 
(a) Schematic of the projection and imaging paths through the dual-channel laparoscope. The LED projector illuminates a passive binary mask, and the pattern is relayed through the illumination channel, while the CMOS camera images the sample through the imaging channel via a folding prism. 
(b) Bench-top implementation showing the LED projector, $L_1$/pinhole relay, prism, CMOS camera, dual-channel endoscope, and laterally positioned Zivid $2^{+}$~M60 depth camera. The Zivid camera is used only for ground-truth depth acquisition during training and is not part of the deployed single-shot sensor. The bottom-left inset shows the inserted binary mask. 
(c) Representative FLIR endoscopic image of the intestinal phantom under projected binary-pattern illumination.}
\label{fig:optical-setup}
\end{figure}

Figure~\ref{fig:optical-setup} shows the optical configuration and bench implementation of the proposed binary-pattern single-shot FPP platform. In the schematic shown in Fig.~\ref{fig:optical-setup}(a), the LED projector illuminates a passive binary mask, and the projected pattern is relayed by lens $L_1$ and a pinhole into the illumination channel of the dual-channel laparoscope (outer diameter $\varnothing = 12$~mm). The imaging channel collects the pattern-distorted sample image and relays it through a folding prism onto a monochrome FLIR camera. Because the illumination and imaging channels share the distal endoscope tip, the projected binary pattern remains aligned with the camera field of view during endoscope motion.

Figure~\ref{fig:optical-setup}(b) shows the bench-top realization of the system. The LED projector, inserted binary mask, $L_1$/pinhole relay, prism, a CMOS camera (FLIR), and dual-channel endoscope are mounted on an optical breadboard. A Zivid $2^{+}$~M60 structured-light depth camera is positioned laterally and off-axis from the endoscope and is used only during data collection to provide independent ground-truth depth. At deployment, the Zivid camera is removed and the endoscope operates as a stand-alone single-shot depth sensor. Figure~\ref{fig:optical-setup}(c) shows a representative FLIR endoscopic image of the intestinal phantom under binary-pattern illumination.

\subsection{Binary-pattern characterization across working distance}
\label{sec:pattern-char}

A practical binary-pattern single-shot system must preserve usable pattern contrast across laparoscopic working distances. We therefore characterized the visibility of the projected binary pattern on an inclined depth-of-field (DOF) target imaged through the endoscope. The distal end of the endoscope was positioned at representative working distances of 7 and 10~cm from the center of the target, near the 25~mm depth marker. For each distance, a rectangular region of interest (ROI) was selected over the projected binary stripes, and the mean grayscale intensity was extracted along the DOF target marker direction.

\begin{figure}[H]
\centering
\includegraphics[width=0.98\linewidth]{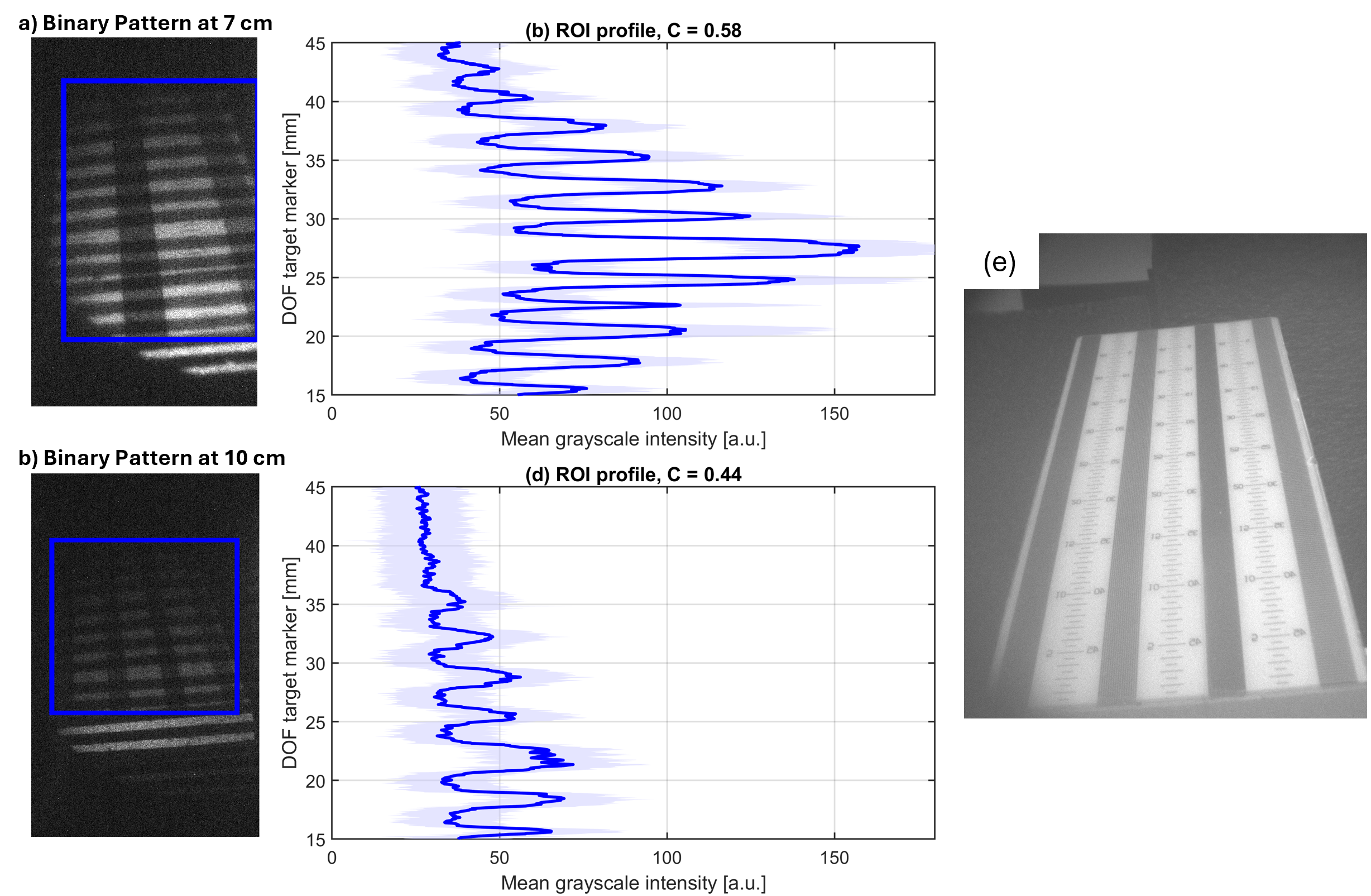}
\caption{Binary-pattern visibility on an inclined depth-of-field target at representative laparoscopic working distances. 
(a,c) Raw SSLE images acquired through the endoscope with the LED-illuminated passive binary mask projected onto the target at 7 and 10~cm working distances, respectively. Blue rectangles indicate the ROIs used for intensity analysis. 
(b,d) Corresponding mean grayscale intensity profiles extracted from the selected ROIs and plotted against the DOF target marker scale. The shaded region indicates the variation across the ROI width, and the solid blue curve indicates the mean profile. The repeated peaks and valleys confirm that the projected binary stripe modulation remains resolvable at both distances, with estimated Michelson contrasts of $C=0.58$ and $C=0.44$. 
(e) Reference endoscopic view of the DOF target used for the working-distance characterization.}
\label{fig:depth-intensity}
\end{figure}

As shown in Fig.~\ref{fig:depth-intensity}, the projected binary stripes remain visually resolvable at both working distances. The ROI-based intensity profiles show repeated peak-to-valley modulation, with estimated Michelson contrasts of $C=0.58$ at 7~cm and $C=0.44$ at 10~cm. Although the mean intensity and contrast decrease with increasing working distance, the binary modulation remains distinguishable, confirming that the LED-illuminated passive mask provides sufficient pattern visibility for single-shot depth reconstruction over laparoscopic-scale working distances.

\subsection{Automated data acquisition and training dataset}
\label{sec:data}

A Zivid-referenced phantom dataset was collected using the proposed LED/binary-mask system shown in Fig.~\ref{fig:optical-setup}. The dataset consists of $722$ paired phantom acquisitions. To ensure repeatable one-to-one pairing between the endoscopic image and the reference depth measurement, an automated collection program was developed to coordinate the CMOS endoscopic camera, the Zivid camera, and the motorized linear stage.

During acquisition, the intestinal phantom was mounted on the linear stage and translated along the viewing direction of the endoscope to sample different working distances and surface poses. At each stage position, the program moved the phantom to the commanded location, allowed the stage to settle, captured a grayscale SSLE image under binary-pattern illumination, and then acquired the corresponding Zivid RGB image, point cloud, and depth map. All files were saved using a common frame index to preserve pairing between the SSLE input image, Zivid color image, Zivid depth map, and acquisition metadata.

Each paired sample consists of a SSLE grayscale image acquired through the laparoscope imaging channel and a dense depth map acquired by the laterally mounted Zivid camera. The CMOS camera image captures the binary-pattern-distorted phantom scene, while the Zivid depth map provides the reference depth measurement. The Zivid depth was subsequently reprojected into the SSLE coordinate frame using the procedure described in Section~\ref{sec:reprojection}. Using an independent depth sensor decouples the training target from the single-shot reconstruction algorithm itself and allows the network to be supervised using an external structured-light depth measurement.

Of the $722$ paired acquisitions, the dataset was divided into training, validation, and held-out test subsets. The training set was used to optimize the network parameters, the validation set was used for model selection and hyperparameter monitoring during training, and the held-out test set was used only for final performance evaluation. All reported quantitative metrics were computed on the held-out test set.

\subsection{Reference-depth generation via cross-camera reprojection}
\label{sec:reprojection}

For every SSLE frame, a dense reference-depth map is obtained from the laterally mounted Zivid structured-light camera and reprojected into the SSLE image frame. A one-time stereo calibration (MATLAB Camera Calibrator Toolbox) between the two cameras yields the intrinsic matrices $K_{\mathrm{SSLE}}$ and $K_{\mathrm{Zivid}}$ together with the rigid transform $(R, \mathbf{t})$ mapping points from the Zivid coordinate frame to the SSLE coordinate frame.

The reprojection proceeds in four steps. First, each valid Zivid pixel $(u_z,v_z)$ with depth $Z_z$ is back-projected into a 3-D point in the Zivid camera frame:
\begin{equation}
\mathbf{P}_z \;=\; Z_z\, K_{\mathrm{Zivid}}^{-1}\,[u_z,\,v_z,\,1]^{\top}.
\label{eq:backproject}
\end{equation}
Second, $\mathbf{P}_z$ is transformed into the SSLE frame by the rigid-body transformation obtained from calibration:
\begin{equation}
\mathbf{P}_f \;=\; R\,\mathbf{P}_z \;+\; \mathbf{t}.
\label{eq:rigid}
\end{equation}
Third, $\mathbf{P}_f=(X_f,Y_f,Z_f)^{\top}$ is projected onto the SSLE image plane via the pinhole model:
\begin{equation}
u_f \;=\; f_{x,f}\,\frac{X_f}{Z_f} + c_{x,f}, \qquad
v_f \;=\; f_{y,f}\,\frac{Y_f}{Z_f} + c_{y,f},
\label{eq:project}
\end{equation}
where $(f_{x,f},f_{y,f},c_{x,f},c_{y,f})$ are the SSLE intrinsics. Fourth, because multiple Zivid rays may project to the same SSLE pixel, a z-buffer is applied that retains the nearest depth:
\begin{equation}
D_{\mathrm{SSLE}}(u_f,v_f) \;=\; \min_{k\,\in\,\mathcal{S}(u_f,v_f)}\, Z_f^{(k)},
\label{eq:zbuffer}
\end{equation}
where $\mathcal{S}(u_f,v_f)$ is the set of Zivid pixels whose projection falls on the SSLE pixel $(u_f,v_f)$. A binary validity mask $M$ is defined to be $1$ where a reprojected Zivid depth exists inside the SSLE image bounds and $0$ elsewhere; the reconstruction loss of Eq.~(\ref{eq:depth-loss}) and the evaluation metrics of Eq.~(\ref{eq:metrics}) are evaluated only at pixels with $M=1$.

The training input is the SSLE grayscale image after the same rotation/flip applied during calibration, denoted $I_{\mathrm{SSLE}}^{\mathrm{proc}}$; a training pair is therefore $(I_{\mathrm{SSLE}}^{\mathrm{proc}},\,D_{\mathrm{SSLE}},\,M)$. Any residual error in $(R,\mathbf{t})$ manifests as a spatial misalignment between $I_{\mathrm{SSLE}}^{\mathrm{proc}}$ and $D_{\mathrm{SSLE}}$ and can thereby introduce an irreducible component of the training loss and evaluation error; we return to this point in Section~\ref{sec:phantom} and in the discussion.

\subsection{Depth-reconstruction network}
\label{sec:network}

The depth-reconstruction network is the algorithmic contribution of this work. The prior design of Zuo et al.\cite{Zuo2025JBO} used two parallel U-Nets: MaskNet for foreground segmentation and DepthNet for depth regression whose outputs were combined by pixel-wise multiplication. Although effective, this parallel design required an explicit masking branch and introduced additional computational complexity. We also observed substantial frame-to-frame variability in inference latency when the pipeline was executed in a real-time loop (Section~\ref{sec:speed}).

\begin{figure}[H]
\centering
\includegraphics[width=0.95\linewidth]{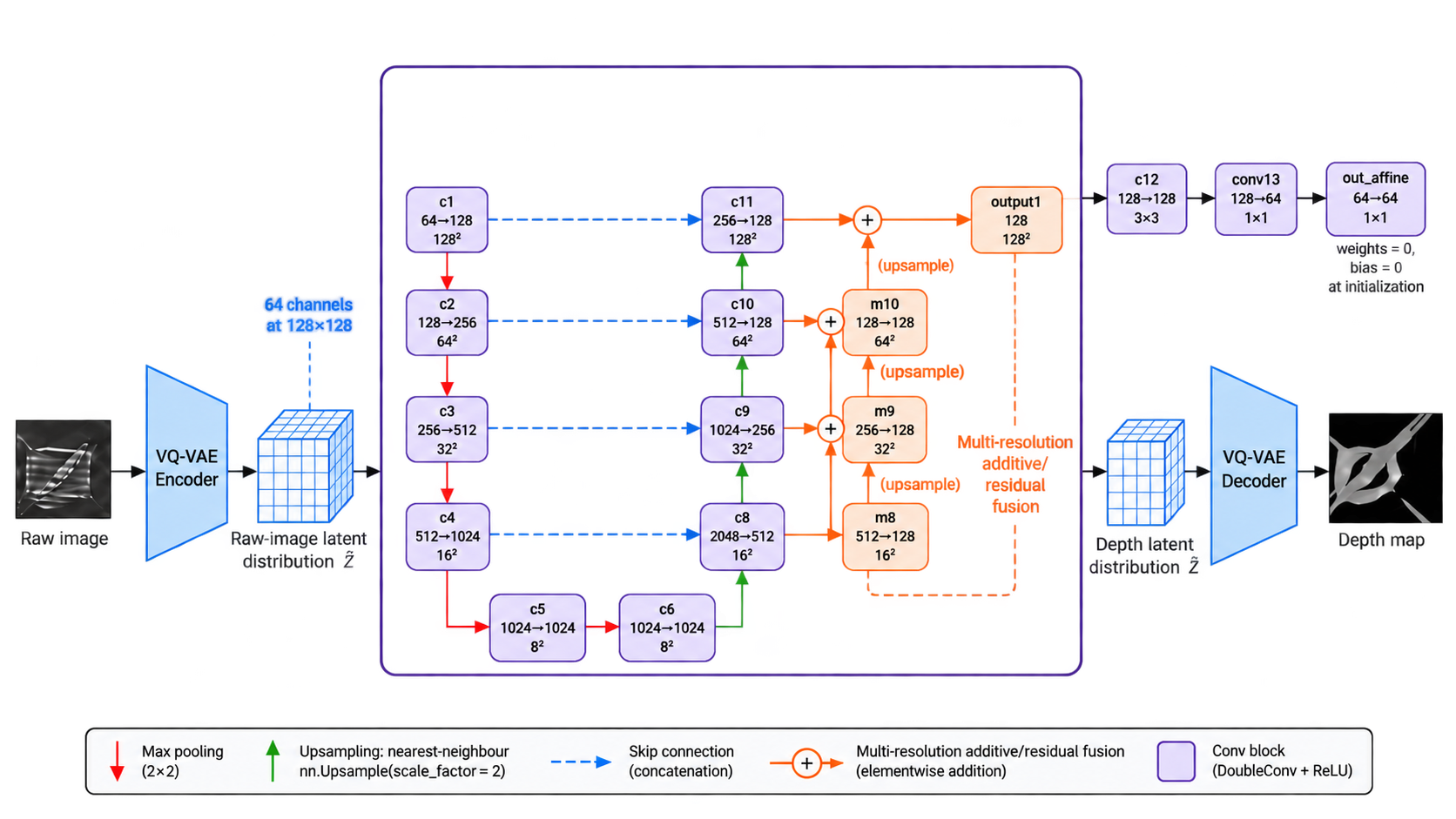}
\caption{Proposed depth-reconstruction network. A VQ-VAE encoder maps the raw pattern-distorted input image to a quantized latent representation $\hat{z}_{\mathrm{in}}$. An U-net with a multi-resolution residual branch(MR U-net) operates in this latent space to predict a latent depth representation $\hat{z}_{\mathrm{depth}}$, which is decoded to produce the final dense depth map. The learned codebook constrains the reconstruction to latent features observed in the training distribution.}
\label{fig:arch}
\end{figure}

To address these limitations, we replaced the previous two-branch design with a single-branch pipeline that incorporates a vector-quantized variational autoencoder (VQ-VAE)\cite{vanDenOord2017VQVAE} as a latent-space prior. As shown in Fig.~\ref{fig:arch}, the pipeline processes a $256 \times 256$ grayscale fringe image $x \in \mathbb{R}^{1 \times 256 \times 256}$ through three sequential components.

\subsubsection*{VQ-VAE image encoder.}
The image encoder maps the fringe image to a discrete latent representation. Its convolutional backbone has base channel width $c=64$ with channel multiplier $[1,\,2]$, yielding one spatial downsampling stage and channel widths of $\{64,\,128\}$ at successive resolutions. Two residual blocks are applied at each resolution level, and a self-attention layer is included at spatial resolution $16{\times}16$. A $1{\times}1$ convolution projects the resulting feature map to the embedding dimension, producing a continuous latent tensor $z_{\mathrm{in}} \in \mathbb{R}^{64 \times 128 \times 128}$. Each spatial position is then independently quantized by nearest-neighbor lookup in a learned codebook of $K_{\mathrm{img}}=512$ entries, each of dimension $d=64$, to produce the discrete latent representation $\hat{z}_{\mathrm{in}} \in \mathbb{R}^{64 \times 128 \times 128}$.

\subsubsection*{Latent U-Net.}
A four-level encoder--decoder U-Net with multi-resolution residual connections (MR U-Net) operates directly on $\hat{z}_{\mathrm{in}}$. The encoder successively halves the spatial resolution through max-pooling and doubles the channel width using paired $3{\times}3$ convolution blocks, producing feature maps at spatial resolutions $128{\times}128$, $64{\times}64$, $32{\times}32$, $16{\times}16$, and $8{\times}8$ (bottleneck) with channel widths $128$, $256$, $512$, $1024$, and $1024$, respectively. The decoder reconstructs resolution via bilinear upsampling with skip connections from corresponding encoder levels. In addition, a multi-level residual branch aggregates decoder features across scales before the final output projection, which maps the decoded features to $\hat{z}_{\mathrm{depth}} \in \mathbb{R}^{64 \times 128 \times 128}$. The base channel width of the U-Net is set to $128$.

\subsubsection*{VQ-VAE depth decoder.}
The depth decoder maps $\hat{z}_{\mathrm{depth}}$ back to pixel space. Its architecture mirrors the image encoder in reverse: base channel width $c=64$, multiplier $[1,\,2]$, two residual blocks per level, and one transposed-convolution upsampling step. The codebook contains $K_{\mathrm{depth}}=1024$ entries of dimension $d=64$. The decoder output is passed through a $\tanh$ activation and linearly rescaled to the training depth range $[0,\,250]$\,mm, producing the final dense depth map $\hat{D} \in \mathbb{R}^{1 \times 256 \times 256}$.

The VQ-VAE serves two purposes in this framework. First, the discrete codebooks constrain the latent representations to the learned distribution of fringe images and depth maps, acting as a structured prior over plausible inputs and outputs. Second, operating in the $64{\times}128{\times}128$ latent space removes the need for explicit foreground segmentation and provides a more compact reconstruction pathway. As shown in Section~\ref{sec:speed}, this design also provides stable frame-to-frame inference timing in a real-time loop.

\subsection{Loss function, training details, and evaluation metrics}
\label{sec:training}

Our loss function consists of a spatial reconstruction loss $\mathcal{L}_{\text{spatial}}$ and a latent-space matching loss $\mathcal{L}_{\text{latent}}$,
\begin{equation}
\mathcal{L}_{\text{total}} =
\mathcal{L}_{\text{spatial}}(D,\hat{D},M)
+ \beta(t)\mathcal{L}_{\text{latent}}(z_d,\hat{z}_d).
\label{eq:total-loss-v2}
\end{equation}
The spatial loss penalizes discrepancies in the final depth-output space over valid Zivid-reprojected pixels using a weighted combination of an $\ell_1$ loss and a structural similarity (SSIM) loss\cite{WangSSIM2004}:
\begin{equation}
\mathcal{L}_{\text{spatial}}(D,\hat{D},M)
=
\alpha \| (D-\hat{D})\otimes M \|_1
+
(1-\alpha)
\left[
1-\mathrm{SSIM}(\tilde{D}\otimes M,\tilde{\hat{D}}\otimes M)
\right],
\label{eq:depth-loss}
\end{equation}
where $\alpha$ balances the two objectives, $M$ is the valid-pixel mask, and $\otimes$ denotes element-wise multiplication. Here, $\tilde{D}$ and $\tilde{\hat{D}}$ denote the reference and estimated depth maps linearly normalized into $[0,1]$ based on the global dataset range.

The latent-space loss is defined as
\begin{equation}
\mathcal{L}_{\text{latent}} = \| z_d-\hat{z}_d \|_1,
\end{equation}
where $z_d$ is the standardized reference-depth latent representation obtained from the pretrained depth VQ-VAE encoder, and $\hat{z}_d$ is the predicted depth latent representation. The weight $\beta(t)$ is controlled by a linear ramp scheduler, which gradually increases from $\beta_{\text{start}}$ to $\beta_{\text{end}}$ during training, encouraging the network to align mid-level representations with the learned quantized structural prior as training stabilizes.

Training was performed in PyTorch\cite{Paszke2019PyTorch} on an NVIDIA A100 GPU. Separate VQ-VAE models were trained for the input-image domain and the reference-depth domain using the taming-transformers framework. Each VQ-VAE was trained with a base learning rate of $5\times10^{-5}$, a batch size of $4$, gradient accumulation over $4$ steps, and the Adam optimizer,\cite{Kingma2015Adam} giving an effective batch size of $16$.

After training the VQ-VAE models, the MR U-Net depth-prediction network was trained for $161$ epochs with a batch size of $4$ using the Adam optimizer and an initial learning rate of $1\times10^{-3}$. A cosine annealing learning-rate schedule was used with $T_{\max}=160$ and a minimum learning rate of $1\times10^{-6}$. Images and depth maps were resized to $256\times256$ for training and evaluation.

We report four evaluation metrics standard in depth regression,\cite{Eigen2014Depth}
\begin{equation}
\mathrm{MAE}=\frac{1}{|V|}\sum_{i\in V}|Y_i-\hat{Y}_i|,
\qquad
\mathrm{AbsRel}=\frac{1}{|V|}\sum_{i\in V}\frac{|Y_i-\hat{Y}_i|}{Y_i},
\end{equation}

\begin{equation}
\mathrm{Acc}_\delta=
\frac{|\{i\in V:\max(Y_i/\hat{Y}_i,\hat{Y}_i/Y_i)<\delta\}|}{|V|},
\label{eq:metrics}
\end{equation}
where $Y=D\otimes M$, $\hat{Y}=\hat{D}\otimes M$, and $V$ is the set of valid pixels with $M=1$, and $\mathrm{Acc}_\delta$ is reported at $\delta=1.1$ and $\delta=1.1^2$. Here, $D$ is the Zivid-reprojected reference-depth map, $\hat{D}$ is the predicted depth map, and $M$ is the Zivid validity mask. Public monocular depth models were evaluated using the same valid-pixel mask and metric definitions. Because these models do not directly predict calibrated endoscopic structured-light depth, they are included as off-the-shelf reference baselines rather than as fully calibrated competitors.

\section{Results}
\label{sec:results}

\subsection{Baseline comparison on the Zivid-referenced dataset}
\label{sec:arch-benchmark}

We evaluated the proposed VQ-VAE + U-Net pipeline against a dual U-Net
MaskNet + DepthNet baseline and two public monocular depth estimation models
on the Zivid-referenced phantom dataset, in which reference depth is obtained
from a laterally mounted Zivid structured-light camera and reprojected into
the SSLE image frame. The $722$ paired SSLE--Zivid acquisitions were divided
into fixed training, validation, and test sets: each model was trained from
scratch on the training set, hyperparameters were selected on the validation
set, and all quantitative results below were computed on the held-out test
set. Metrics were evaluated only over valid pixels defined by the Zivid
reprojection mask, to exclude background regions without reliable Zivid depth.

\begin{figure}[H]
\centering
\includegraphics[width=0.98\linewidth]{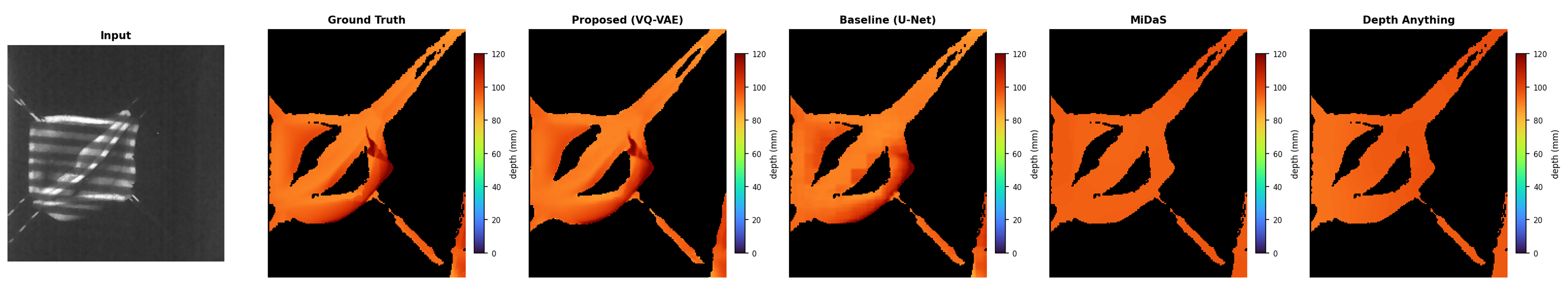}
\caption{Qualitative depth-reconstruction comparison on a representative
held-out sample from the Zivid-referenced phantom dataset. The raw SSLE input
image, reprojected Zivid reference depth, proposed VQ-VAE + U-Net result
(shown without applying the Zivid validity mask), dual U-Net MaskNet + DepthNet
baseline, and public monocular depth model outputs are shown. The proposed
model preserves the dominant phantom geometry and produces a depth distribution
visually consistent with the reprojected Zivid reference depth even though no
segmentation branch is used. The largest errors occur near the edges of the
validity mask, around thin structures, and in areas of local SSLE--Zivid
reprojection mismatch.}
\label{fig:model-comparison}
\end{figure}

Figure~\ref{fig:model-comparison} shows a representative qualitative
comparison. The raw output of the proposed model, shown without applying the
Zivid validity mask, already closely resembles the Zivid-reprojected reference
depth in both spatial extent and depth distribution, indicating that the
network learned an implicit representation of the foreground phantom region
from the structured-light image appearance. In contrast, the MaskNet +
DepthNet baseline relies on an explicit mask-prediction branch whose output is
combined with the predicted depth map to form the final reconstruction. The
public monocular depth models recover coarse object structure but do not
preserve the calibrated endoscopic depth distribution as accurately.

\begin{table}[H]
\centering
\caption{Depth-reconstruction comparison on the Zivid-referenced phantom
dataset, evaluated on valid masked pixels only. Public monocular depth models
are included as off-the-shelf reference baselines.}
\label{tab:zivid-comparison}
\resizebox{\textwidth}{!}{%
\begin{tabular}{lcccc}
\toprule
Model & MAE (mm) $\downarrow$ & AbsRel $\downarrow$ 
      & Acc $(1.1)$ $\uparrow$ & Acc $(1.1^2)$ $\uparrow$ \\
\midrule
Proposed (VQ-VAE + U-Net)
  & $3.70$ & $0.0326$ & $0.962$ & $0.970$ \\
Baseline (dual U-Net: MaskNet + DepthNet)
  & $4.30$ & $0.0560$ & $0.965$ & $0.970$ \\
MiDaS DPT-Large
  & $8.33$ & $0.0739$ & $0.755$ & $0.929$ \\
Depth Anything
  & $7.80$ & $0.0691$ & $0.772$ & $0.942$ \\
\bottomrule
\end{tabular}%
}
\end{table}

As summarized in Table~\ref{tab:zivid-comparison}, the proposed VQ-VAE +
U-Net model achieved an MAE of $3.70$~mm and AbsRel of $0.0326$, compared
with $4.30$~mm and $0.0560$ for the MaskNet + DepthNet baseline, while the
baseline retained a slight advantage in threshold accuracy at $\delta = 1.1$
($0.965$ vs.\ $0.962$). Both values are higher than the $2.28$~mm MAE reported for the MaskNet +
DepthNet architecture in its original setting \cite{Zuo2025JBO}, where the
supervisory depth was generated by the multi-shot FPP algorithm itself; here, both models are instead supervised and evaluated
against independently measured, cross-camera-reprojected Zivid depth, which
introduces an additional error component discussed in
Section~\ref{sec:Discussion}. The two designs exhibit different error
characteristics: the proposed model lowers the average reconstruction error,
whereas the mask-guided baseline keeps marginally more pixels within the
relative-error threshold. Beyond the quantitative comparison, the proposed
model eliminates the dedicated mask-prediction branch during depth generation;
qualitative inspection shows that the raw unmasked predictions already
approximate the valid phantom support and suppress background depth
hallucination. Both task-specific models substantially outperformed the
off-the-shelf monocular depth baselines across MAE, AbsRel, and threshold
accuracy, underscoring the value of task-specific supervision for calibrated
depth recovery in this endoscopic structured-light setting.

\subsection{Inference speed and frame-rate stability}
\label{sec:speed}

\begin{figure}[H]
\centering
\includegraphics[width=0.7\linewidth]{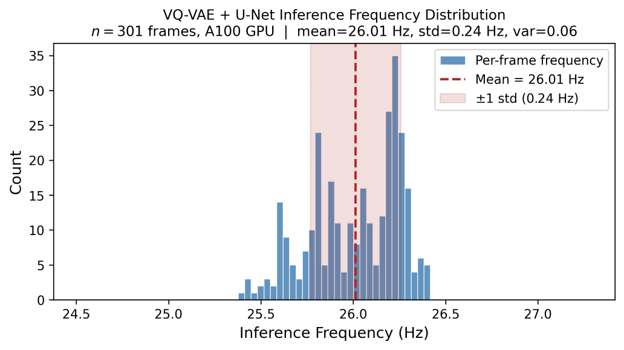}
\caption{Inference-frequency distribution for the proposed VQ-VAE + U-Net pipeline over $301$ consecutive frames on an NVIDIA A100 GPU. The pipeline achieves a mean inference frequency of $26.0$~Hz, demonstrating stable video-rate inference performance.}
\label{fig:speed}
\end{figure}

Figure~\ref{fig:speed} reports the per-frame inference frequency distribution for the proposed VQ-VAE + U-Net pipeline over $301$ consecutive frames on an NVIDIA A100 GPU. Only model inference time was recorded; data loading, visualization, and file I/O were excluded from the timing measurement.

The pipeline achieved a mean inference frequency of $26.0$~Hz, corresponding to a mean per-frame latency of approximately $38.4$~ms. This frame rate is within the video-rate regime commonly used for intraoperative image-guidance and laparoscopic video-processing systems, where approximately 30~fps is often treated as a practical real-time target~\cite{Ronaghi2015}. Because the frame-to-frame latency is predictable, a surgical display or guidance system can be designed around a fixed processing budget rather than accommodating large frame-rate fluctuations.

\subsection{Phantom depth reconstruction with Zivid-referenced depth}
\label{sec:phantom}

\begin{figure}[H]
\centering
\includegraphics[width=0.84\linewidth]{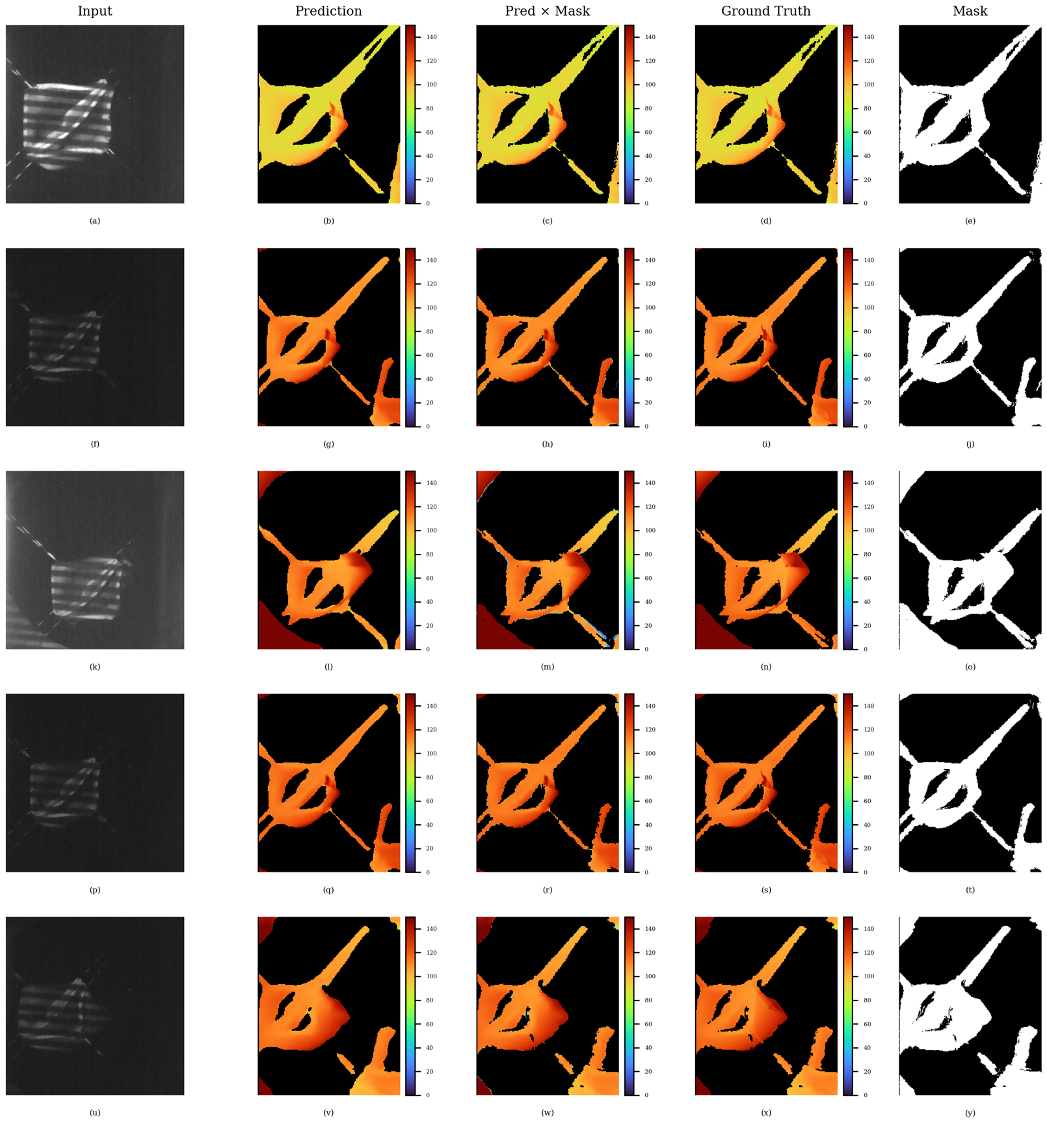}
\caption{Phantom depth reconstruction with Zivid-referenced depth. Five representative held-out examples are shown from the binary-pattern SSLE--Zivid phantom dataset. Columns show the raw SSLE input image, VQ-VAE + U-Net depth prediction, prediction multiplied by the Zivid validity mask, reprojected Zivid reference depth, and the corresponding validity mask used for evaluation. Across the five displayed examples, the MAE, AbsRel, and accuracy at $\delta=1.1$ are $3.01$~mm, $0.0264$, and $0.972$, respectively. The network recovers the overall phantom geometry across different test views, with the largest deviations at mask boundaries, thin structures, and locally misaligned regions.}
\label{fig:phantom-results}
\end{figure}

We further evaluated the proposed pipeline qualitatively on representative held-out samples from the Zivid-referenced phantom dataset, in which the reference depth for each SSLE frame is supplied by a laterally mounted Zivid camera and reprojected into the SSLE image frame. Figure~\ref{fig:phantom-results} shows five representative test examples. For each case, the raw SSLE input, VQ-VAE + U-Net prediction, masked prediction, reprojected Zivid reference depth, and Zivid validity mask are shown side by side. The network recovers the dominant phantom geometry across different views, including the folded lobes and connecting ridge structures.

For the five representative examples shown in Fig.~\ref{fig:phantom-results}, the proposed pipeline achieved an overall MAE of $3.01$~mm, AbsRel of $0.0264$, and accuracy at $\delta=1.1$ of $0.972$. These values are reported only for the displayed examples and are intended to complement the held-out test-set metrics reported in Table~\ref{tab:zivid-comparison}. The predicted depth maps are visually coherent within the phantom interior and remain consistent across changes in phantom pose and field-of-view location. The unmasked predictions preserve continuous phantom structure before applying the Zivid validity mask, whereas the masked predictions inherit the hard boundaries and local omissions of the reprojected Zivid mask. Thus, the validity mask is used to define reliable evaluation regions rather than as a required component of the proposed depth-generation pipeline.

The remaining discrepancies concentrate near the edges of the validity mask, around thin structures, and in regions where the Zivid-referenced depth contains reprojection artifacts or local misalignment relative to the SSLE image. Such errors are expected: pixel-level mismatch between the SSLE input and the reprojected Zivid reference depth sets a floor on both the training signal and the measured error that the network cannot reduce. This is consistent with prior work showing that RGB-D/depth-camera calibration, depth accuracy, and registration errors can affect reconstructed depth measurements and image-depth alignment \cite{Khoshelham2012,Lachat2015}.

\section{Discussion}
\label{sec:Discussion}
The results above support three main conclusions. First, the proposed VQ-VAE + U-Net pipeline achieved competitive depth reconstruction on the Zivid-referenced phantom dataset with no dedicated segmentation stage in the depth-generation path. On the held-out test set from the $722$ paired SSLE--Zivid acquisitions, the proposed model achieved an MAE of $3.70$~mm, AbsRel of $0.0326$, accuracy at $\delta=1.1$ of $0.962$, and accuracy at $\delta=1.1^2$ of $0.970$. The latent-space reconstruction framework can therefore recover calibrated depth from single-shot binary-pattern images using Zivid-reprojected reference depth for supervision. Relative to the dual U-Net MaskNet + DepthNet baseline, the proposed model trades a marginal loss in threshold accuracy for a lower average error, and its raw unmasked predictions approximate the valid phantom support on their own, which is evident that the network learned the foreground structure implicitly. The Zivid validity mask therefore serves only to define reliable evaluation regions, not as a component of the deployed pipeline.

Second, the remaining reconstruction errors are partly attributable to the Zivid-referenced supervision strategy. The reference-depth maps depend on cross-camera reprojection from the laterally mounted Zivid camera into the SSLE image frame. Residual SSLE--Zivid calibration mismatch, mask-boundary uncertainty, and phantom pose or field-of-view variation can introduce spatial mismatch between the SSLE input image and the reprojected Zivid reference depth. Because the network cannot compensate for supervision that is itself misaligned, this mismatch contributes a floor to both the training loss and the evaluation metrics, most visibly near validity-mask boundaries, thin structures, and local depth discontinuities. The same supervision difference explains why both the proposed model and the reimplemented MaskNet + DepthNet baseline report higher MAE here than the ${\sim}2.28$~mm previously obtained when the same baseline architecture was trained and evaluated against FPP-derived depth \cite{Zuo2025JBO}. In that setting, the ground truth was produced by the same optical pathway and camera as the input image, so the network was not penalized for calibration or reprojection error. Under Zivid-referenced supervision, any residual SSLE--Zivid extrinsic error contributes to the reported metrics for every model evaluated, so the absolute MAE values are not directly comparable across the two supervision protocols. This is consistent with known limitations of RGB-D/depth-camera calibration and registration \cite{Khoshelham2012,Lachat2015}.

Third, the pipeline sustained a stable mean inference frequency of $26.0$~Hz (${\approx}38.4$~ms per frame) over $301$ consecutive frames on an NVIDIA A100 GPU, meeting the practical requirements of video-rate surgical guidance described in Section~\ref{sec:speed}. Further profiling of the encoder, U-Net, decoder, and data-transfer components will be needed to identify the main computational bottlenecks and improve the frame rate.

Several limitations should be stated explicitly. The current experiments were performed on tissue phantoms, so validation on \textit{ex vivo} and \textit{in vivo} tissue remains necessary to assess the effects of specular highlights, blood, and tissue deformation\cite{MaierHein2014TMI}. The Zivid-referenced depth maps are also limited by the accuracy of the SSLE--Zivid extrinsic calibration and cross-camera reprojection. Future work will therefore focus on improved calibration, tissue experiments, and evaluation under more realistic surgical imaging conditions.

\section{Conclusion}

We have described a compact LED-illuminated binary-pattern single-shot structured-light platform and a VQ-VAE-regularized depth reconstruction network for real-time robotic laparoscopy. The hardware eliminates the DMD projector, projector--camera synchronization, and associated driver electronics, reducing the projection module to a passive mask, an LED source, and relay optics that fit within the illumination channel of a standard dual-channel laparoscope. On a $722$-image Zivid-referenced phantom dataset, the proposed network achieved an MAE of $3.70$~mm and accuracy at $\delta=1.1$ of $0.962$ using a fixed train/validation/test split, with no explicit segmentation stage in the depth-generation path. The pipeline also maintained stable video-rate inference at approximately $26.0$~Hz. Reference-depth supervision was decoupled from the legacy FPP reconstruction algorithm by using an independent Zivid structured-light camera and reprojecting its depth into the endoscopic camera frame. The current SSLE--Zivid-referenced results remain limited by cross-camera calibration error and phantom-only validation, but they establish the feasibility of synchronization-free, real-time, learning-based single-shot depth sensing for robotic surgical guidance.

\subsection*{Disclosures}
The authors declare no conflicts of interest. OpenAI ChatGPT was used solely for grammar and language refinement. All scientific content, ideas, analyses, and references were developed by the authors and carefully reviewed to ensure accuracy and integrity.

\subsection*{Code, Data, and Materials Availability}
Code and phantom data are available from the corresponding author on reasonable request.

\subsection*{Acknowledgments}
This work was supported in part by the U.S. National Institutes of Health under award 1R01EY032127. The authors thank the Rockfish High Performance Computing facility at Johns Hopkins University for compute resources, and members of the Intelligent Optical Imaging and Vision (iOIV) and Laboratory for Computational Sensing and Robotics (LCSR) groups at JHU for helpful discussions.

\bibliographystyle{spiejour}
\bibliography{references}

\vspace{1em}
\noindent\textbf{Wayne Wonseok Rodgers} is a PhD student in the Department of Electrical and Computer Engineering at Johns Hopkins University, working in Prof. Jin U. Kang's group.
\par\medskip
\noindent\textbf{Xiangyi Le} was a Master's student with the Laboratory for Computational Sensing and Robotics at Johns Hopkins University during the performance of this work.
\par\medskip
\noindent\textbf{Seonghoon Jang} is an undergraduate student in the Department of Mechanical Engineering, pursuing a combined BS/MS degree program.
\par\medskip
\noindent Biographies of the others are not available.

\end{document}